\documentclass[11pt]{article}
\usepackage{amsmath,amssymb,amsfonts}

\newtheorem{theorem}{Theorem}

\def\R{{\mathbb R}}
\def\C{{\mathbb C}}

\def\H{{\mathbb H}}

\def\Re{{\mathrm{Re}\, }}

\def\tilde{\widetilde}

\begin{document}

\title{On a formation of singularities of solutions to soliton equations represented by $L,A,B$-triples
\thanks{The work was supported by RSCF (grant 19-11-00044-P).}}
\author{Iskander A. TAIMANOV
\thanks{Novosibirsk State University, 630090 Novosibirsk, and Sobolev Institute of Mathematics, 630090 Novosibirsk, Russia;
e-mail: taimanov@math.nsc.ru}}
\date{}
\maketitle

\begin{abstract}
We discuss the mechanism of formation of singularities of solutions to the Novikov--Veselov, modified Novikov--Veselov, and Davey--Stewartson II (DSII) equations obtained by the Moutard type 
transformations. These equations admit the $L,A,B$-triple presentation, the generalization of the $L,A$-pairs for 2+1-soliton equations. We relate the blow-up of solutions to the non-conservation of the zero level of discrete spectrum of the  $L$-operator. We also present a class of exact solutions, of the DSII system, which depend on two functional parameters, and show that all possible singularities of solutions to DSII equation obtained by the Moutard transformation are indeterminancies, i.e.,
points when approaching which in different spatial directions the solution has different limits. 
\end{abstract}

One of important problems in theory of nonlinear equations concerns a formation of singularities of solutions.
This includes a problem of blowing up of solutions. It is interesting to understand possible mechanisms of that and conditions which lead to appearance of singularities during the evolution.

Here we would like to discuss such a problem for soliton equations which are represented by so-called ``L,A,B''-triples \cite{Manakov}.

We recall the notion of the $L,A,B$-triple in Section 1.

In Section 2 we discuss the examples of blowing up solutions for the Novikov--Veselov, modified Novikov--Veselov, and Davey--Stewartson II equations
and point out that for these examples at the terminal time the zero level of the discrete spectrum changes
and that relates to the appearances of singularities of solutions. 

In Section 3 we describe the Moutard transformation for the Davey--Stewartson II equation and discuss the types of singularities which can be constructed by the Moutard transformation for solutions of the modified Novikov--Veselov and Davey--Stewartson II equations. We show that these singularities are only indeterminancies, i.e., points when approaching which in different spatial directions the solution has different limits.

In Section 4 we make a couple of remarks on questions which to our opinion it would be interesting to answer.

\section{$L,A,B$-triples}
\subsection{$L,A$-pairs}

The inverse scattering method for integrating one-dimensional soliton equations is based on the Lax pair representation of the equation which is its representation in the form
$$
L_t = [L,A]
$$
where $L$ and $A$ are ordinary differential operators.
The most famous and initial example is given by the
the Korteweg--de Vries (KdV) equation
$$
u_t = 6uu_x - u_{xxx}
$$
for which
$$
L = -\frac{d^2}{dx^2} + u(x,t)
$$
and $A$ is an ordinary differential operator of third order.
These evolutions preserve the spectrum of the operator $L$ deforming the eigenfunctions as
$$
\psi_t + A\psi = 0, \ \ \ \mbox{where $L\psi = E\psi$}.
$$
The inverse scattering method is based on the conservation of the spectrum and on
the linearization of the soliton equation on the spectral data. For instance, for the KdV equation we assume that
$$
\int_{-\infty}^{\infty} |u(x)| (1+|x|)\,dx < \infty
$$
(the Faddeev condition \cite{Faddeev, Marchenko}). Then the potential $u(x)$ is uniquely recon\-struct\-ed
via the Gelfand--Levitan--Marchenko equation from the spectral data which is
$$
r(k) \ \mbox{(the reflection coefficient)}, \ k \in \R \setminus \{0\},
$$
$$
\lambda_j = -\beta_j^2 < 0 \  \mbox{with $\beta_j >0$ (the bound states)}, \ \
c_j \ \mbox{(the normed coefficients)}.
$$
The bound states form the discrete spectrum which is finite and the normed coefficients
$$
c_j = \left( \int_{-\infty}^{+\infty} \psi_j^2 dx\right)^{-1}
$$
are
determined by the corresponding eigenfunctions
$$
L \psi_j = \lambda_j \psi_j
$$
normalized by the condition
$$
\psi_j \sim e^{-\beta_j x} \ \ \mbox{as $x \to +\infty$}.
$$
Under the stronger condition $\int_{-\infty}^{\infty} |u(x)| (1+x^2)\,dx < \infty$, the inverse scattering problem was investigated in \cite{DT} where, in particular, it was showed that under this condition the scattering matrix $S(k)$ is continuously extended onto the whole line.

On the spectral data the KdV flow is linear:
$$
\frac{\partial r(k,t)}{\partial t} = 8ik^3 r(k,t), \ \ \ \frac{dc_j}{dt} = -8\lambda_j c_j.
$$

The well-posedness of the Cauchy problem for the KdV equation was well-studied and established in wide functional classes \cite{KPV,KV}. Similar problems were studied for the sin-Gordon equation and the nonlinear Schr\"odinger equation which are also integrated by using the inverse scattering method.

\subsection{Two-dimensional soliton equations represented by $L,A,B$-triples}

$L,A,B$-triples were introduced by Manakov in the middle of 1970s as 2D equations presented in the form
$$
L_t = [L,A] + BL
$$
where $L,A$, and $B$ are partial differential operators.

Manakov came to these systems because he did not succeed to construct nontrivial $L,A$-pairs for two-dimensional systems for which $L$ is a two-dimensional version of some classical operator from mathematical physics. It appeared to be possible only if  the potential of $L$ is a sum of one-dimensional potentials which depend on $x$ and $y$. He argued that the spectrum of $L$ is not preserved except its zero level
$$
L\psi = 0
$$
which evolves as in the case of Lax pairs:
$$
\psi_t + A\psi = 0.
$$
Of course, this argument is formal and it is as follows:
if $L\psi = 0$, then
$$
(L\psi)_t = L_t \psi + L \psi_t = LA \psi - AL\psi +BL\psi +L \psi_t  =
L(\psi_t + A\psi).
$$
We see that, if $\psi_t+A\psi = 0$, then  $(L\psi)_t = 0$, and therefore $L\psi =0$.
The important point that

\begin{itemize}
\item
{\sl this argument assumes that a solution of the soliton and a solution of the equation
$\psi_t + A\psi = 0$ are regular}.
\end{itemize}

Otherwise it may fail and we show below that that may happen.

Let us recall the most known and well studied equations which admit the $L,A,B$-triple representation.

1) The Novikov--Veselov (NV) equation:
$$
U_t = U_{zzz} + U_{\bar{z}\bar{z}\bar{z}}  + 3(VU)_z +
3 (\bar{V}U)_{\bar{z}} =0, \ \ \
V_{\bar{z}} = U_z,
$$
with
$$
L = \partial \bar{\partial} + U,
$$
$U=U(z,\bar{z},t), U = \bar{U}, z \in \C$.

It was introduced by Novikov and Veselov \cite{VN} in the framework of theory of two-dimensional 
Schr\"odinger operators finite-gap on one energy level \cite{DKN}. In one-dimensional limit, $U=U(x)$, it reduces to the KdV equation.

This equation is initial for a hierarchy of soliton equations for which $L$ is a two-dimensional Schr\"odinger operator.

2) The modified Novikov--Veselov(mNV)  equation:
$$
U_t = \big(U_{zzz} + 3U_z V + \frac{3}{2}UV_z \big) + \big(U_{\bar{z}\bar{z}\bar{z}} + 3U_{\bar{z}}\bar{V} + \frac{3}{2} U\bar{V}_{\bar{z}}\big),
$$
$$
V_{\bar{z}} = (U^2)_z
$$
where
\begin{equation}
\label{dirac}
L = \left(
\begin{array}{cc}
0 & \partial \\
-\bar{\partial} & 0
\end{array}
\right) + \left(
\begin{array}{cc}
U & 0 \\
0 & U
\end{array}
\right),
\ \ \
U = \bar{U}.
\end{equation}

This equation was introduced in the late 1980s by Bogdanov \cite{Bogdanov}.
In one-dimensional limit, $U=U(x)$, it reduces to the modified Korteweg--de Vries equation.
It is the initial equation for the hierarchy associated with two-dimensional Dirac operators (\ref{dirac}) with real-valued potentials.

3) The Davey--Stewartson II (DSII) equation:
\begin{equation}
\label{dsii}
U_t = i(U_{zz}+U_{\bar{z}\bar{z}} + (V+\bar{V})U),  \ \ \
V_{\bar{z}} = 2(|U|^2)_z,
\end{equation}
where
\begin{equation}
\label{diracc}
L = \left(
\begin{array}{cc}
0 & \partial \\
-\bar{\partial} & 0
\end{array}
\right) + \left(
\begin{array}{cc}
U & 0 \\
0 & \bar{U}
\end{array}
\right).
\end{equation}
This equation was introduced in \cite{DS} and describes certain surface waves.
In soliton theory it was extended to the hierarchy associated with the two-dimensional Dirac operators (\ref{diracc}) with complex-valued potentials.

\section{Formation of singularities of solutions and the discrete spectrum of $L$-operators}

By using the Darboux--Moutard transformations (see, for instance, \cite{MS,TT10}) blowing up solutions
were constructed for the Novikov--Veselov \cite{TT08}, modified Novikov--Veselov \cite{Taimanov2015}, and
Davey--Stewartson II \cite{Taimanov2021} equations.

These examples have the following features:

\begin{itemize}
\item
the initial data for all such solutions are fast decaying and belong to the $L_2$-space (of functions for the NV equations, and of vector-functions for the mNV and DSII equations);

\item
the functional
$$
\int_{\R^2} U dx dy
$$
is a first integral of the NV equation. The blowing up solutions are regular and fast decaying up to the
certain time $T_{sing}$ when it becomes singular and stays singular afterwards;

\item
the functional
\begin{equation}
\label{willmore}
W = \int_{\R^2}  |U|^2 dx dy
\end{equation}
is the first integral of the mNV and DSII equations. The blowing up solutions are regular except one point
$(x_0,y_0,T_{sing})$ in the 2+1-dimensional space. For $t< T_{sing}$ and $t>T_{sing}$  the functional $W$ has the same values however for $t=T_{sing}$ it is defined and has a different value.
\end{itemize}

The construction of  examples for the mNV and DSII equations is based on the spinor (Weierstrass) representation of surfaces in the three- and four-spaces and their soliton deformations \cite{Taimanov06}.
Therewith the functional $W$ is proportional to the Willmore functional and the solutions are constructed from soliton deformations of branched Willmore spheres. This implies that in these examples
the values of $W$ are proportional to $\pi$.

What sounds interesting for all examples is the behavior of the discrete spectrum.

For the case of the NV equation the Schr\"odinger operator with the initial potential has the negative and the zero parts of the discrete spectrum. The zero part has the multiplicity two. In \cite{AT}   we presented the numerical evidence that the negative part of the discrete spectrum tends to $-\infty$ as
$t \to T_{sing}$.

For the cases of the mNV and DSII equations the corresponding Dirac operators
have eigenfunctions with the zero eigenvalue.

We expose these examples following  \cite{TT08,Taimanov2015,Taimanov2021}.
In addition we supply the formulas for blowing up eigenfunctions on the zero energy level
which can be extracted from these papers but were not explicitly presented in them.

\begin{theorem}
1) \, Given
$$
F(x,y,t) = 2(x^4 + y^4) + \frac{8}{3}(x^3 + y^3) + 4x^2 y^2 +20 - 8t,
$$
for
$$
t <  T_{sing} = \frac{29}{12};
$$
the functions
$$
U = 2\partial \bar{\partial} F, \ \ V = 2\partial^2 F
$$
satisfy the Novikov--Veselov equation
and
$$
L \psi = (\partial \bar{\partial} + U) \psi = 0, \ \ \ \psi = \frac{xy}{F};
$$

2) \, Given
$$
a = i(x^2-y^2), \ \ b = -y\left(\frac{y^2}{3}-x^2-1\right) - i(x\left(1+y^2 - \frac{x^2}{3}\right) + C-t), \ \ C \in \R,
$$
for
$$
t \neq T_{sing} = C
$$
the function
$$
U = i\frac{a(|z|^2-1) + \bar{b}\bar{z}-bz}{|a|^2+|b|^2}
$$
satisfies the modified Novikov--Veselov equation and
$$
L \psi = \left[\left(
\begin{array}{cc}
0 & \partial \\
-\bar{\partial} & 0
\end{array}
\right) + \left(
\begin{array}{cc}
U & 0 \\
0 & U
\end{array}
\right)\right]\psi = 0
$$
where
$$
\psi = \frac{1}{|a|^2+|b|^2} \left(\begin{array}{c} \bar{a}z- \bar{b} \\ \bar{a}+\bar{b}\bar{z}\end{array}\right)
$$
and  the polynomial $|a|^2+|b|^2$ vanishes if and only if $C=t$.
Moreover
$$
\int_{\R^2} U^2 \, dx dy =
\begin{cases} 3 \pi & \mbox{for $t \neq C$}, \\
2\pi & \mbox{for $t=C$}.
\end{cases}
$$

3) For
$$
t \neq T_{sing} \neq C \in \R
$$
the function
$$
U = i\frac{z^2 - 2i(t-C)}{|z|^2 + |z^2 +2i(t-C)|^2}
$$
satisfies the Davey--Stewartson II equation  and
$$
L \psi = \left[\left(
\begin{array}{cc}
0 & \partial \\
-\bar{\partial} & 0
\end{array}
\right) + \left(
\begin{array}{cc}
U & 0 \\
0 & \bar{U}
\end{array}
\right)\right]\psi = 0
$$
where
$$
\psi = \frac{1}{|z|^2+|z^2 + 2i(t-C)|^2} \left(\begin{array}{c} \bar{z} \\
-iz^2 + 2(t-C)\end{array}\right).
$$
Moreover
$$
\int_{\R^2} |U|^2 \, dx dy =
\begin{cases} 2 \pi & \mbox{for $t \neq C$}, \\
\pi & \mbox{for $t=C$}.
\end{cases}
$$
\end{theorem}

We expose only the simplest examples but by the methods proposed in \cite{TT08,Taimanov2015,Taimanov2021} we may construct more such solutions depending on functional parameters. Theorem 1 leads to the following observation:

\begin{itemize}
\item
{\sl for solutions given by Theorem 1 and similar ones constructed by the same methods
the evolution does not preserve the zero level of the discrete spectrum because as
$t \to T_{sing}$ certain eigenfunctions on the zero energy level are becoming singular, i.e., blow up.}
\end{itemize}

\section{The Moutard transformation for the DSII equa\-tion}

Construction of exact solutions of the DS II equation attended a lot of attention and was discussed in many papers. We do not discuss them here because we are interested only in formation of singularities and, in particular, the new mechanism of that found in \cite{Taimanov2021} and based on the soliton deformation of surfaces in the four-space via the DSII hierarchy \cite{Konopelchenko,TaimanovDS}.
 
The DSII equation is the compatibility condition for the system
$$
L \Psi =0 , \ \ \ \partial_t \Psi = A\Psi
$$
where
$$
A = i\left(\begin{array}{cc} -\partial^2 - V & \bar{U}\bar{\partial} - \bar{U}_{\bar{z}} \\
U\partial - U_z & \bar{\partial}^2 + \bar{V} \end{array}\right)
$$
and
$$
\Psi =
\left(\begin{array}{cc} \psi_1 & -\bar{\psi}_2 \\
\psi_2 & \bar{\psi}_1 \end{array} \right).
$$
It is easy to check that $L\psi = 0$  if and only if $L\Psi=0$ where
$\psi = \left(\begin{array}{c} \psi_1 \\ \psi_2 \end{array}\right)$.
The DSII equation is also the compatibility condition for the system
$$
L^\vee \Phi = 0, \ \ \Phi_t = A^\vee \Phi,
$$
with
$$
L^\vee = \left(\begin{array}{cc} 0 & \partial \\
-\bar{\partial} & 0 \end{array}\right) +
\left(\begin{array}{cc} \bar{U} & 0 \\
0 & U \end{array}\right),\ \
A^\vee = - i\left(\begin{array}{cc} -\partial^2 - V & U\bar{\partial} - U_{\bar{z}} \\
\bar{U}\partial - \bar{U}_z & \bar{\partial}^2 + \bar{V} \end{array}\right),
$$
and
$$
\Phi = \left(\begin{array}{cc} \varphi_1 & -\bar{\varphi}_2 \\
\varphi_2 & \bar{\varphi}_1 \end{array} \right).
$$

To describe the Moutard transformation we correspond to every pair $\Psi$ and $\Phi$ of functions with values
in $u(2) = \H$, the Lie algebra of $U(2)$ identified with quaternions, i.e., of the form
$$
\left(\begin{array}{cc} \lambda & -\bar{\mu} \\ \mu & \bar{\lambda} \end{array}\right) \in \H, \ \
\lambda,\mu \in \C,
$$
the $1$-form
$$
\omega(\Phi,\Psi) =
-\frac{i}{2}\left(\Phi^\top \sigma_3 \Psi + \Phi^\top \Psi\right) dz -
\frac{i}{2}\left(\Phi^\top \sigma_3 \Psi - \Phi^\top \Psi\right) d\bar{z},
$$
where  $X \to X^\top$ is the transposition of $X$ and
$\sigma_3= \left(\begin{array}{cc} 1 & 0 \\ 0 & -1 \end{array}\right)$ is one of the Pauli matrices.
If $L\Psi = L^\vee \Phi=0$, then the forms $\omega(\Phi,\Psi)$ and $\omega(\Psi,\Phi)$ are closed
and there is defined an $\H$-valued function
\begin{equation}
\label{sigma}
S(\Phi,\Psi)(z,\bar{z}) = \Gamma \int \omega(\Phi,\Psi)
\end{equation}
where $\Gamma = \left(\begin{array}{cc} 0 & 1 \\ -1 & 0 \end{array}\right)= i\sigma_2$.
Then let us defined another $\H$-valued function
\begin{equation}
\label{k}
K(\Phi,\Psi) =  \Psi S^{-1}(\Phi,\Psi)\Gamma \Phi^\top\Gamma^{-1} =
\left(\begin{array}{cc} i\bar{W} & a \\ -\bar{a} & -iW \end{array}\right).
\end{equation}

In \cite{MT} it was showed that

if $L\Psi_0 = L^\vee\Phi_0$ =0, then
for every pair $\Psi$ and $\Phi$ of solutions of the same equations
the $\H$-valued functions
$$
\widetilde{\Psi} =  \Psi - \Psi_0 S^{-1}(\Phi_0,\Psi_0) S(\Phi_0,\Psi), \
\widetilde{\Phi} =  \Phi - \Phi_0 S^{-1}(\Psi_0,\Phi_0) S(\Psi_0,\Phi)
$$
satisfy the Dirac equation
$$
\widetilde{L}\widetilde{\Psi} = 0, \ \ \ \ \widetilde{L}^\vee \widetilde{\Phi} = 0
$$
for operators with the potential
\begin{equation}
\label{newpotential}
\widetilde{U} = U + W
\end{equation}
where $W$ is defined by (\ref{k}) for $K(\Phi_0,\Psi_0)$.
Here it is assumed that
$$
\Gamma S^{-1}(\Phi_0,\Psi_0)\Gamma = (S^{-1}(\Psi_0,\Phi_0))^\top
$$
which is achieved by choosing integration constants in definitions of $S$-matrices (\ref{sigma}).

This transformation is a generalization of the Moutard transformation derived in \cite{DLS} for the equation
$L\Psi=0$ with a real-valued potential $U$ to which it reduces for $U=\bar{U}$ and $\Psi_0=\Phi_0$ and for solutions of the mNV equation. 
We rewrote this transformation to interpret it geometrically. The $S$-matrix defines an immersion of surface with a conformal parameter $z$ into the imaginary quaternions $\R^3 = su(2) \subset u(2)$,
the operator $L$ is the Dirac operator coming in its spinor representation defined by $\Psi_0$ \cite{Taimanov06}, and $\widetilde{L}$ and $\widetilde{\Psi}_0$ give the spinor representation of the inversed surface $S^{-1}$.

This geometrical interpretation helped to find singular solutions to the mNV equation \cite{Taimanov2015}
(see also (2) of Theorem 1): a singularity of $\widetilde{U}$ appears when the surface $S(x,y,t)$ passes through the origin and hence the corresponding point is mapped into $\infty$.

In \cite{Taimanov2021} the transformation (\ref{newpotential}) was extended to a transformations of solutions of the DSII equation. For that it needs to pick up solutions $\Psi_0$ and $\Phi_0$ of the linear systems
$L\Psi_0 = (\partial_t-A)\Psi_0 = 0$ and $L^\vee\Phi_0 = (\partial_t - A^\vee)\Phi_0=0$ and use the same formulas with $S$ replaced by
$$
S(\Phi,\Psi)(z,\bar{z},t) = \Gamma \int \omega(\Phi,\Psi) + \Gamma \int \omega_1(\Phi,\Psi)
$$
with
$$
\omega_1(\Phi,\Psi) =
\left(\left[
\Phi^\top_z \left(\begin{array}{cc} 1 & 0 \\ 0 & 0 \end{array}\right) +
\Phi^\top_{\bar{z}}\left(\begin{array}{cc} 0 & 0 \\ 0 & 1 \end{array}\right)
\right] \Psi
\right.
$$
$$
\left.
- \Phi^\top\left[
\left(\begin{array}{cc} 1 & 0 \\ 0 & 0 \end{array}\right)\Psi_z +
\left(\begin{array}{cc} 0 & 0 \\ 0 & 1 \end{array}\right)\Psi_{\bar{z}}\right]\right) dt.
$$
Therewith the function $\tilde{U}$ satisfies the DSII equation
$$
\tilde{U}_t = i(\tilde{U}_{zz}+\tilde{U}_{\bar{z}\bar{z}} + 2(\tilde{V}+\bar{\tilde{V}})\tilde{U}), \ \ \
\tilde{V}_{\bar{z}} = (|\tilde{U}|^2)_z,
$$
with
\begin{equation}
\label{newv}
\tilde{V} = V + 2ia_z
\end{equation}
and $a$ given by (\ref{k}).

The geometrical interpretation is similar: $\widetilde{U}$ is the potential of the spinor representation of
the inversed surface $S^{-1} \subset \R^4 \cup \infty$ and the singular points correspond to the solutions of the equation $S=0$.

The simplest way to construct new solutions is to start with $U=0$. In \cite{Taimanov2021}
it was proved that if $f(z,t)$ is a holomorphic in $z$ function such that
$$
\frac{\partial f}{\partial t} = i\frac{\partial^2 f}{\partial z^2} = if^{\prime\prime},
$$
then
$$
U = \frac{i(zf^\prime - f)}{|z|^2 + |f|^2}, \ \
V = 2ia_z \ \ \mbox{with $a = -\frac{i(\bar{z} + f^\prime) \bar{f}}{|z|^2 + |f|^2}$}
$$
meet the DSII equation. It is obtained by the Moutard transformation determined by
$$
\Psi_0 = \left(\begin{array}{cc} 0 & -1 \\ 1 & 0 \end{array}\right), \ \
\Phi_0 = \left(\begin{array}{cc} f^\prime & i \\ i & \bar{f}^\prime \end{array}\right).
$$
In particular, for
$$
f = z^2 + 2it -2i C
$$
we have the solution given in Theorem 1 which has a singularity at the origin for $t=C$:
\begin{equation}
\label{indet1}
U \sim i e^{2i \phi} \ \ \ \mbox{as $r  \to 0$}, \ \ \mbox{where $z = r e^{i\phi}$}.
\end{equation}
We remark that if $C$ is not real then we have a solution which is regular everywhere and is fast decaying as
$r \to \infty$. Therefore such a singularity is unstable with respect to small variations of the initial data.

Since exact solutions of the DSII equation attract a lot of attention, we present 
a new class which depends on two functional parameters:

\begin{theorem}
Let $f(z,t)$ and $g(z,t)$ be holomorphic in $z$ functions such that
$$
\frac{\partial f}{\partial t} = i \frac{\partial^2 f}{\partial z^2}, \ \ \ 
\frac{\partial g}{\partial t} = - i \frac{\partial^2 g}{\partial z^2}.
$$
Then the function
$$
U = i \bar{g}^\prime \frac{f^\prime g - h}{|g|^2 + |h|^2},
$$
where $h^\prime = f^\prime g^\prime$, satisfies the DSII equation.
\end{theorem}

This theorem is derived by applying the Moutard transformation to the pair
$$
\Psi_0 = \left(\begin{array}{cc} g^\prime & 0 \\  0 &\bar{g}^\prime \end{array}\right), \ \
\Phi_0 = \left(\begin{array}{cc} f^\prime & i \\ i & \bar{f}^\prime \end{array}\right).
$$
For instance, for 
$$
f(z,t) = z^2 + 2it, \ \ g(z,t) = z^3 - 6itz
$$
we derive 
$$
U(z,t) = i(3\bar{z}^2+6it) \frac{\frac{1}{2}z^4 - 6zitz^2}{|z^3- 6itz|^2 + |\frac{1}{2}z^4-6itz^2|^2}
$$
which for $t=0$ has the singularity of the form
$$
U \sim \frac{3}{2}i e^{2i\phi} \ \ \mbox{as $r \to 0$}.
$$
We have $U = O(r^{-2})$ as $r \to \infty$. By increasing the power of the polynomial $g$ in $z$, we may achieve  faster decay of $U$ at the infinity.

There appears the question: can we obtain by the Moutard transformation more complicated singularities
than indeterminancies at certain points?

In \cite{TT08} the desired non-singular initial data were obtained by using iterations of the Moutard transformation. We can also write down formulas for iterations starting at $U=0$ for our case however 
we do not achieve more complicated singularities.

Let derive from (\ref{k}) the formula for $W$: 
\begin{equation}
\label{k1}
W = -i \frac{\alpha \bar{\psi}_1 \varphi_1 - \beta \psi_2 \varphi_1 + \bar{\alpha}\psi_2 \bar{\varphi}_2 +
\bar{\beta}\bar{\psi}_1\bar{\varphi}_2}{|\alpha|^2+|\beta|^2}
\end{equation}
with
$$
S(\Phi_0,\Psi_0) = \left(\begin{array}{cc} \alpha & -\bar{\beta} \\ \beta & \bar{\alpha} \end{array}\right).
$$
From these formulas we conclude that

\begin{itemize}
\item
{\sl If $U$ is smooth as well as $\Psi$ and $\Phi$, then $\tilde{U} = U+W$ gains singularities exactly at points where $\alpha = \beta=0$, i.e., where $S=0$. In this case both numerator and denominator in (\ref{k1}) vanish and the indeterminacies of $U$ appear at these points.}
\end{itemize}  

The same conclusion for the same reasons is valid for solutions of the mNV equation constructed by the Moutard transformation. 

\section{Remarks}

1) In \cite{Ozawa} Ozawa constructed the first known blowing up solution for the DSII equation.
Making the change of variables
$$
X = 2y, \ \ Y = 2x,
$$
we rewrite the DSII equation in the form
$$
iU_t -U_{XX} + U_{YY} = -4|U|^2U + 8\varphi_X U,
$$
$$
\Delta\varphi = \frac{\partial^2 \varphi}{\partial X^2}
+ \frac{\partial^2 \varphi}{\partial Y^2} = \frac{\partial}{\partial X} |U|^2,
$$
where $\Re V = 2|U|^2 - 4 \varphi_X$and $\varphi_X = \frac{\partial \varphi}{\partial X}$.
Ozawa had taken the initial data in the form
$$
U(X,Y,0) = \frac{e^{-ib(4a)^{-1}(X^2 - Y^2)}}{a(1+ ((X/a)^2+(Y/a)^2)/2)}
$$
and showed that for the constants $a$ and $b$ such that
$ab<0$, we have
$$
\| U\|^2 \to 2 \pi\cdot \delta \ \mbox{as $t \to T = -a/b$}
$$
in ${\cal S}^\prime$ where $\|U\|^2 = \int_{\R^2}|U|^2\, dx\,dy$ and $\delta$ is the Dirac distribution centered at the origin. This solution is extended to a regular one for $T > -a/b$. For the survey of blowing up solutions of the DSII equations we refer to \cite[\S 5]{KS}.

It would be interesting to find the spectral data for the Ozawa potentials and understand are they related to the blow-up mechanism. In difference with solutions given by the Moutard transformation the Ozawa blow-up is expected to be stable: the numerical evidence for that was presented in \cite{KSt}.

2) In \cite{Croke}  all initial data for the Novikov--Veselov equation were splitted into three natural classes: subcritical, critical, and supercritical, due to the behaviour of the Schr\"odinger operator on the zero energy level. It was conjectured that the Novikov--Veselov equation has a global solution for critical and subcritical initial data, but its solution may blow up in finite time for supercritical initial data.  Is that related to 
the non-conservation of the zero energy level of the discrete spectrum?


\begin{thebibliography}{99}

\bibitem{AT}
Adilkhanov, A.N., and Taimanov, I.A.:
On numerical study of the discrete spectrum of a two-dimensional Schr\"odinger operator with soliton potential.
Comm. Nonlin. Sci. Numer. Simulation {\bf 42} (2017), 83--92.

\bibitem{Bogdanov}
Bogdanov L.V.:
Veselov--Novikov equation as a natural two-dimensional
generalization of the Korteweg--de Vries equation.
Theor. Math. Phys. {\bf 70} (1987), 219--223.

\bibitem{Croke}
Croke, R., Mueller, J.L., Music, M., Perry, P., Siltanen, S., and Stahel, A. 
The Novikov-Veselov equation: theory and computation.
In: {\it Nonlinear Wave Equations: Analytic and Computational Tehcnique}, Contemporary Mathematics {\bf 635}, pp. 25--70, AMS, Providence, 2015. 

\bibitem{DS}
Davey, A., and Stewartson, K.:
On three--dimensional packets of surface waves.
Proc. R. Soc. Lond. A {\bf 38} (1974), 101--110.

\bibitem{DT}
Deift, P., and Trubowitz, E.:
Inverse scattering on the line.
Comm. Pure and Appl. Math. {\bf 32} (1979), 121--251.

\bibitem{DKN}
Dubrovin B.A., Krichever I.M., Novikov S.P.:
The Schr\"odinger equation in a periodic field and Riemann surfaces.
Soviet Math. Dokl. {\bf 17} (1976), 947--952.

\bibitem{Faddeev}
Faddeev, L.D.:
Properties of the S-matrix of the one-dimensional Schr\"odinger operator, Amer. Math. Soc., Transl., II Ser. 65, 1967, pp. 139--166.

\bibitem{KPV}
Kenig, C., Ponce, G., and Vega, L.:
Well-posedness of the initial value problem for the Korteweg-de Vries equation.
J. Amer. Math. Soc. {\bf 4} (1991), 323--347.

\bibitem{KV}
Killip, R., and Visan, M.:
KdV is well-posed in $H^{-1}$.
Ann. of Math. (2) {\bf 190}(1) (2019), 249--305.

\bibitem{KS}
Klein, C., and Saut, J.-C.:
IST versus PDE: a comparative study.
In: Hamiltonian partial differential equations and applications. Fields Inst. Commun.
{\bf 75}.Fields Inst. Res. Math. Sci., Toronto, Ontarion, 2015, 383--449.

\bibitem{KSt}
Klein, C., and Stoilov, N.:
Numerical study of blow-up mechanisms for Davey--Stewartson II systems.
Stud. Appl. Math. {\bf 141}:1 (2018), 89--112.

\bibitem{Konopelchenko}
Konopelchenko, B.G.: 
Weierstrass representations for surfaces in 4D
spaces and their integrable deformations via DS hierarchy. 
Annals of Global Anal. and Geom. {\bf 16} (2000), 6--74.

\bibitem{Manakov}
Manakov, S.V.:
The method of the inverse scattering problem and twodimensional evolution equations.
Uspekhi Mat. Nauk {\bf 31}:5 (1976), 245--246. (Russian)

\bibitem{Marchenko}
Marchenko, V.A.:
Sturm--Liouville operators and their applications.
Naukova Dumka, Kiev, 1977. (Russian)

\bibitem{MT}
Matuev, R.M., and Taimanov, I.A.:
The Moutard transformation of two-dimensional Dirac operators and the conformal geometry of surfaces in four-space. Math. Notes {\bf 100}:6 (2016), 835--846.

\bibitem{MS}
Matveev, V.B., and Salle, V.A.:
Darboux transformations and solitons.
Springer Series in Nonlinear Dynamics. Springer-Verlag, Berlin, 1991.

\bibitem{Ozawa}
Ozawa, T.:
Exact blow-up solutions to the Cauchy problem for the Davey--Stewartson systems.
Proc. Roy. Soc. London Ser. A {\bf 436}:1897 (1992), 345--349.

\bibitem{TaimanovDS}
Taimanov, I.A.: 
Surfaces in the four-space and the Davey--Stewartson equations. 
J. of Geometry and Physics {\bf 56} (2006), 1235--1256.

\bibitem{Taimanov06}
Taimanov, I.A.:
Two-dimensional Dirac operator and surface theory.
Russian Math. Surveys {\bf 61}:1 (2006), 79--159.

\bibitem{Taimanov2015}
Taimanov, I.A.:
Blowing up solutions of the modified Novikov--Veselov equation and minimal surfaces.
Theor. Math. Phys. {\bf 182} (2015), 173--181.

\bibitem{Taimanov2021}
Taimanov, I.A.:
The Moutard transformation for the Davey--Stewartson II equation and its geometrical meaning.
Math. Notes {\bf 110}:5 (2021), 754--766.

\bibitem{TT08}
Taimanov, I.A., and Tsarev, I.A.:
Two-dimensional rational solitons and their blow-up via the Moutard transformation.
Theor. Math. Phys. {\bf 157} (2008), 1525--1541.

\bibitem{TT10}
Taimanov, I.A., and Tsarev, S.P.:
On the Moutard transformation and its applications to spectral theory and soliton equations.
J. of Math. Sciences {\bf 170} (2010), 371--387.

\bibitem{VN}
Veselov, A.P, and Novikov, S.P.:
Finite-zone two-dimensional potential Schr\"odinger operators. Explicit formulas and evolution equations.
Soviet Math. Dokl. {\bf 30} (1984), 588--591.


\bibitem{DLS}
Yu, D.L., Liu, Q.P., Wang, S.K.:
Darboux transformation for the modified Veselov--Novikov equation.
J. of Physics A {\bf 35} (2001), 3779--3785.

\end{thebibliography}
\end{document}